\documentclass[prb,twocolumn,showpacs,aps,superscriptaddress,floatfix]{revtex4-2}
\usepackage{amsmath}
\usepackage{amssymb}
\usepackage{amsfonts}
\usepackage{bm}
\usepackage[dvips]{graphicx}
\usepackage{color}
\usepackage{ulem}
\usepackage[unicode, pdftex]{hyperref}
\usepackage{verbatim}% Long comments
\begin{document}
\title{Strain induced spin vortex and Majorana Kramer's pairs in doped topological insulators with nematic superconductivity}
\author{R.S. Akzyanov}
\affiliation{Dukhov Research Institute of Automatics, Moscow, 127055 Russia}
\affiliation{Moscow Institute of Physics and Technology, Dolgoprudny,
    Moscow Region, 141700 Russia}
\affiliation{Institute for Theoretical and Applied Electrodynamics, Russian
    Academy of Sciences, Moscow, 125412 Russia}
\affiliation{P.N. Lebedev Physical Institute, Russian Academy of Sciences, Moscow 119991, Russia}

\author{A.L. Rakhmanov}
\affiliation{Dukhov Research Institute of Automatics, Moscow, 127055 Russia}
\affiliation{Institute for Theoretical and Applied Electrodynamics, Russian
    Academy of Sciences, Moscow, 125412 Russia}

\begin{abstract}
Using the Ginzburg-Landau approach we show that the strain of the nematic superconductor can generate a specific (nematic) vorticity. In the case of doped topological insulators that vorticity forms a spin vortex. We find two types of topologically different spin vortices that either enhance (type I) or suppress (type II) superconductivity far from the vortex core. We apply Bogoliubov-de Gennes equations to study electronic states in the nematic superconductor with spin vortices. We find that in the case of the vortex of type I, zero-energy states are localized near the vortex core. These states can be identified as Majorana Kramer's pairs. In the case of the vortex of type II, there are no localized zero-energy states. Thus, we establish a non-trivial connection between the strain and Majorana fermions in the doped topological insulators with nematic superconductivity.
\end{abstract}
\maketitle

\section {Introduction}

Nematic superconductivity in doped topological insulators attracts a great attention nowadays~\cite{Yonezawa2018,Cho2020,Frohlich2020,Schmidt2020,Das2020,Kuntsevich2020,Akzyanov2020, Akzyanov2020a}. In these systems, the superconducting order parameter is a time-reversal invariant, which corresponds to $E_u$ representation that breaks inversion symmetry and couples electrons with the same spin projections but from different orbitals~\cite{Fu2009,Fu2014}. The NMR measurements confirmed the triplet nature of the nematic topological superconductivity in doped topological insulators~\cite{Matano2016}.

Exotic quasi-particles with non-Abelian statistics such as Majorana fermions can exist in the topological superconductors~\cite{Fu2008,Qi2011}. The Majorana fermions can be localized on various types of topological defects~\cite{Teo2010}. One way to induce Majorana fermions is to generate vorticity in the mass term. For example, the Majorana fermions can be localized in the cores of Abrikosov vortices
~\cite{Volovik1999,Ivanov2001,Fu2008,Akzyanov2015,Akzyanov2016}. If the time-reversal symmetry is present, then, the Majorana fermions arise as Kramer's pairs~\cite{Teo2010,Chiu2016}. 

Superconducting order in the doped topological insulators belongs to DIII symmetry class~\cite{Schnyder2008}. An analog of the nematic superconductor of the class DIII is the superfluid B phase in the $^3$He~\cite{Volovik_book}. An interesting property of such a phase of helium is a possible realization of the spin vortex that preserves the time-reversal symmetry. The spin vortices in the B phase of the $^3$He have been observed experimentally~\cite{Korhonen1993}. 

The spin vortices in the context of superconductivity have been briefly discussed for $(p_x+ip_y)_\uparrow (p_x-ip_y)_\downarrow$ superconductors~\cite{Chiu2016}. The spin vortex (referred to as nematic vortex) was studied in a superconductor with nematic order parameter in Ref.~\onlinecite{Wu2017}. A single-orbital Hamiltonian with a quadratic dispersion and $k$-dependent order parameter was considered. It has been argued that the spin vortex brings Majorana Kramer's pairs into the system that form a Majorana flat band.

A distinct feature of the nematic superconductivity is a strong coupling of the superconductivity with strain~\cite{Venderbos2016}. In particular, the strain is responsible for a two-fold symmetry of the second critical field that has been observed in the experiments~\cite{Kuntsevich2018,Kuntsevich2019}. The strain can be either spontaneous or external~\cite{Akzyanov2020}. We show that the applied centrosymmetric strain can generate spin vortices in doped topological insulators with nematic superconductivity.

We assume that a local force is applied to a sample of the doped topological insulator, which has a form of a disc. The force generates a centrosymmetric strain that couples with the superconductivity and forms a (nematic) spin vorticity. Depending on the sample properties two types of topologically different spin vortices can exist. Such spin vortices have a normal core. We solve Bogoliubov-de Gennes (BdG) equations and show that one type of the spin vortices localizes the Majoarana Kramer's pairs. Near the core of the spin vortex of another type there are no localized zero-energy states.

\section{Ginzburg-Landau approach}

Ginzburg-Landau (GL) free energy of the $E_u$ topological superconductor with $D_{3d}$ crystal symmetry can be written as~\cite{Fu2014}
\begin{eqnarray}\label{F0}
\nonumber
F_0 &=& A(|\Delta_1|^2+|\Delta_2|^2)+B_1\left(|\Delta_1|^2+|\Delta_2|^2\right)^2\\
&+&B_2|\Delta^*_1\Delta_2-\Delta_1\Delta_2^*|^2.
\end{eqnarray}
Here $\vec{\Delta}=(\Delta_1,\Delta_2)$ is the vector order parameter, $A_1\propto T-T_c<0$ and $B_1>0$ are the GL coefficients. We suppose that $B_2$ is positive, which corresponds to the nematic superconductivity with a real order parameter $\vec{\Delta}=\Delta(\cos{\alpha},\sin{\alpha})$. Vector $\vec{n}=(\cos{\alpha},\sin{\alpha})$ shows nematicity direction. The free energy~\eqref{F0} is degenerate with respect to $\alpha$. The nematicity direction can be fixed by the strain~\cite{Fu2014,Venderbos2016}. 

We assume that the sample is deformed by some local external force and the corresponding strain tensor has components $u_{xx}$, $u_{yy}$, and $u_{xy}$, which depend on the coordinate $\vec{r}$. The strain tensor couples with the superconducting order parameter. This coupling is described by an additional term in the GL free energy~\cite{Fu2014,Venderbos2016}     
\begin{eqnarray}
\nonumber\label{Fu}
F_u&=&g_N(u_{xx}-u_{yy})(|\Delta_1|^2-|\Delta_2|^2)\\
&+&2g_Nu_{xy}(\Delta_1^*\Delta_2+\Delta_1\Delta_2^*),    
\end{eqnarray}
where $g_N$ is a coupling coefficient. 
The order parameter becomes coordinate-dependent and, in general, we have to take into account corresponding gradient terms in the GL free energy. However, deformation and superconductivity are different phenomena with their spatial scales. The scale of the superconductivity is the effective coherence length, $\xi_{\textrm{eff}}$. It is a microscopic value, while the strain characteristic scale $l_u$ is a macroscopic value of the order of the sample sizes. Thus, it is reasonable to suppose that $l_u\gg\xi_{\textrm{eff}}$. In this case, away from the center of the vortex $r \gg \xi_{\textrm{eff}}$, we can neglect the gradient terms and assume that $\vec{\Delta}$ depends on the coordinate parametrically, that is, $\vec{\Delta}(\vec{r})=\vec{\Delta}[u_{ik}(\vec{r})]$ and the order parameter can be found from minimization of $F_0+F_u$ with respect to $\vec{\Delta}$. To obtain a correct behavior of the order parameter near the center of the vortex, $r\sim \xi_{\textrm{eff}}$, we should take into account the gradient terms in the GL functional. This procedure is performed in Appendix A. We show that the spin vortex has a normal core with the size $\sim \xi_{\textrm{eff}}$ similar to the Abrikosov vortices~\cite{Abrikosov1957}. This normal core can be considered as a topological defect. The values of the coherent length for the vortex of type I, $\xi_I$, and type II, $\xi_{II}$, are different.    

We suppose that the force, and, hence, the strain, has a rotational symmetry and the strain tensor components can be written in the cylindrical coordinates $(r,\varphi,z)$ as (see Ref.~\onlinecite{LandauLifshtiz7} and Appendix A)
\begin{eqnarray}\label{strain}
\nonumber
u_{xx}-u_{yy}=u(r,z)\cos{(2\varphi)},\\
2u_{xy}=u(r,z)\sin{(2\varphi)},
\end{eqnarray}
where $u(r,z)$ depends on the applied force, sample sizes, and boundary conditions. 

\begin{figure}[b]
\includegraphics[width=1.0\linewidth]{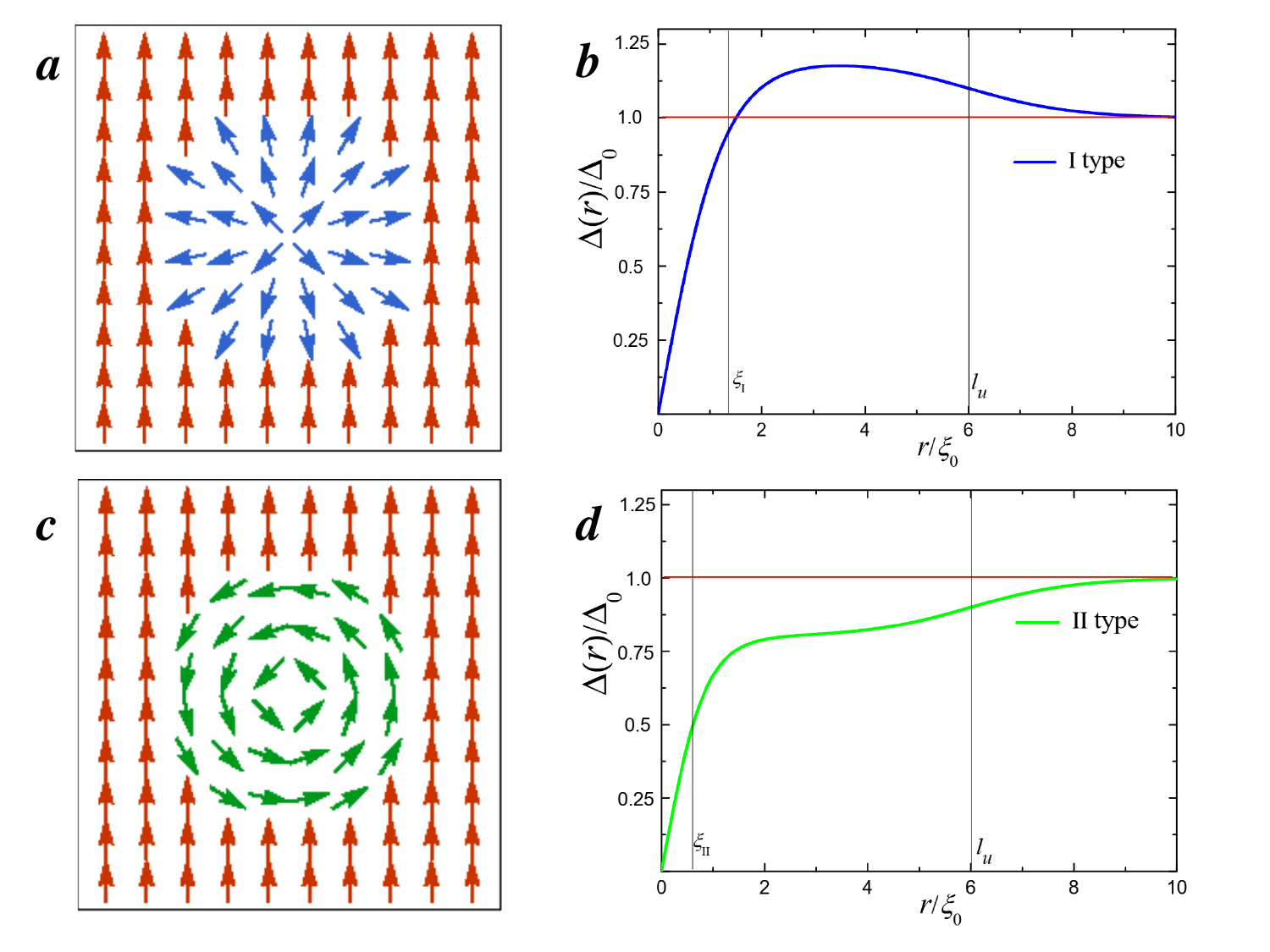}
\caption{ A schematic picture of the nematicity direction $\vec{n}$ ({\it a}) and the function $\Delta(r)/\Delta_0$ ({\it b}) for the vortex of type I. In panels {\it c} and {\it d} the same for the vortex of type II. Here $\xi_I$, $\xi_{II}$, and $\xi_0$ are effective coherence lengths for the vortex of type I, for the vortex of type II, and for the undeformed sample, respectively [formulas for $\xi_i$ are presented in Appendix A, Eqs.~\eqref{GLeta_dim}]. The size of the vortex core is of the order of the corresponding coherence length, which is a microscopic value. The size of the spin vortex is of the order of macroscopic scale $l_u$ of the strain. We assume that for $r>l_u$ the nematicity direction (1,0) is fixed.}
%%%%%%%%%%%%%%%%%%%%%%%%%%%%%%%%%%%%%%%%%%%%%%%%%%%%%%%%%%
\label{Fig1}
%%%%%%%%%%%%%%%%%%%%%%%%%%%%%%%%%%%%%%%%%%%%%%%%%%%%%%%%%%%
\end{figure}

After substitution of expressions for $\vec{\Delta}$ and $u_{ik}$ in Eqs.~\eqref{F0} and \eqref{Fu} we obtain 
\begin{eqnarray}\label{GL_tot}
F_0+F_u=A\Delta^2+B_1\Delta^4+g_Nu\Delta^2\cos{[2(\alpha-\varphi)]}.
\end{eqnarray}
When $g_Nu(r,z)>0$, the minimization of $F_0+F_u$ by $\alpha$ gives 
\begin{equation}\label{alpha01}
\alpha=\varphi+\pi\left(n+\frac{1}{2}\right).    
\end{equation}
When $g_Nu(r,z)<0$, the minimum of $F_0+F_u$ attains if  
\begin{equation}\label{alpha02}
\alpha=\varphi+\pi n.   
\end{equation}
Here $n$ is an integer or zero. The value of $\Delta(r,z)$ is obtained from minimization of the GL free energy with respect to $\Delta$. Taking into account Eqs.~\eqref{alpha01} and~\eqref{alpha02}, we derive 
\begin{equation}\label{eta_r}
\Delta(r,z)=\sqrt{\frac{-A+g_Nu(r,z)}{2B_1}}.   
\end{equation}
Thus, the external force not only affects the value of the order parameter but also forms a vortex in the nematicity $\alpha \propto \varphi$. We have two types of vorticity depending on the sign of $g_Nu(r)$. If $g_Nu(r,z)>0$ (see Eq.~\eqref{alpha01}), we will call a corresponding solution as a spin vortex of type I and
\begin{equation}\label{alpha1}
\vec{\Delta}_I=\Delta(r,z)(\cos{\varphi},\sin{\varphi}).   
\end{equation}
In the case $g_Nu(r,z)<0$ (see Eq.~\eqref{alpha02}), we have a spin vortex of type II:
\begin{equation}\label{alpha2}
\vec{\Delta}_{II}=\Delta(r,z)(-\sin{\varphi},\cos{\varphi}),   
\end{equation}
The superconducting order parameter away from the vortex core is either enhanced (type I) or reduced (type II). A schematic picture of the local nematicity direction and order parameter behavior for these spin vortices are shown in Fig.~\ref{Fig1}. The spin vortices are topologically different, the vector field of the spin vortex of type I looks like a ``hedgehog", while the nematicity vector field for the spin vortex of type II has a form of a ``curl". We can calculate the winding number of the nematicity vector  $\vec{n}(\mathbf{r})$ around the vortex core 
$$P=\oint_C \vec{n}\cdot{\bf dr}/2\pi,$$
where ${\bf dr}=(dx,dy)$ and $C$ is the closed contour around the vortex core with unit radius. In the case of the spin vortex of type I winding number vanishes, $P=0$, while for the spin vortex of type II the winding number is non-zero, $P=1$. Further, we show that different topology of the vortices results in different quasiparticle spectra.

\section{Bogoliubov-de Gennes equations}\label{Section_BdG}

Now we seek zero-energy solutions of the BdG equations assuming that $l_u \rightarrow +\infty$ (for more details see also Appendix B). For the doped topological insulators these equations can be presented as~\cite{Fu2010,Yip2013,Fu2014,Venderbos2016a}
\begin{eqnarray}\label{BdG}
H_{\textrm{BdG}}(\mathbf{k})=H_0(\mathbf{k})\tau_z+\vec{\Delta}\tau_x,
\end{eqnarray}
where single-electron Hamiltonian $H_0$ is
\begin{eqnarray}\label{H0}
H_0(\mathbf{k})\!=\! -\mu \!+\! m\sigma_z\!+\!v\sigma_x(s_xk_y\!-\!s_yk_x)\!+\!v_zk_z\sigma_y.\,\,\,\,
\end{eqnarray}
Here ${\mathbf \sigma}$, ${\mathbf s}$, and ${\mathbf \tau}$ are the Pauli matrices acting in orbital, spin, and electron-hole spaces, respectively, the superconducting order parameter is $\vec{\Delta}=\Delta(r)\sigma_y{\mathbf s}\cdot{\vec{n}}$ (the symmetry of the order parameter corresponds to $E_u$ pairing~\cite{Fu2010,Yip2013,Fu2014,Venderbos2016a}), 
${\mathbf k}$ is the momentum, $\mu$ is the chemical potential, $m$ is a single electron gap, and $v$ and $v_z$ are the in-plane and transverse Fermi velocities. According to the GL consideration, we choose $\vec{n}$ as $\vec{n}=[\cos{(\varphi+\nu \pi/2)},\sin{(\varphi+\nu \pi/2)}]$, where $\nu=0$ corresponds to the spin vortex of type I, Eq.~\eqref{alpha1} and $\nu=1$ corresponds to the vortex of type II,  Eq.~\eqref{alpha2}. The strain induces a so-called pseudomagnetic field in the system with Hamiltonian~\eqref{H0}, however, this field is negligible in the doped Bi$_2$Se$_3$ materials~\cite{Brems2018}.

The spin vortex can be induced in Hamiltonian~\eqref{BdG} by the transformation~\cite{Volovik_book,Chiu2016}
\begin{eqnarray}
e^{-is_z[\varphi+(\nu-1) \pi/4]}\Delta \sigma_y s_x\tau_x  e^{is_z[\varphi+(\nu-1) \pi/4]},
\end{eqnarray}
 The vortex generates vorticity in the spin space ${\mathbf s}$, while the Abrikosov vortex generates vorticity in the mass space ${\mathbf \tau}$~\cite{Volovik_book}.

We consider here the states with $k_z=0$ and rewrite Eq.~\eqref{BdG} in the coordinate space substituting $k_{x(y)}=-i\nabla_{x(y)}$. In the polar coordinates the Hamiltonian reads
    \begin{eqnarray}\label{Bdg1}
 &&H_{\textrm{BdG}}=-\mu \tau_z+m\sigma_z\tau_z+iv \sigma_x s_x\tau_z\times\\
    \nonumber
 && \left[e^{i(\varphi\!+\!\pi/2)s_z}\nabla_r\!-\!\frac{1}r e^{i\varphi s_z}\nabla_{\varphi} \right]
    \!\!+\!\Delta \sigma_y s_x \tau_x e^{i(\varphi\!+\!\nu \pi/2) s_z}\,.
    \end{eqnarray}
We are interested in the solutions of the BdG equations with zero energy: $H_{\textrm{BdG}}\Psi=0$, where eight component spinor is $\Psi=(f_{1\uparrow},f_{1\downarrow},f_{2\uparrow},f_{2\downarrow},h_{1\uparrow},h_{1\downarrow},h_{2\uparrow},h_{2\downarrow})^T$. Here 1 and 2 are orbital indices, $\uparrow$ and $\downarrow$ are spin projections, and $f$ and $h$ represent electron or hole states. We will seek such a solution in the form
\begin{eqnarray}\label{major1}
\Psi(r,\varphi)=\exp{[i(l-s_z/2)\varphi]}\frac{\psi(r)}{\sqrt{r}},
\end{eqnarray}
where $l$ is the orbital number. The wave function is single-valued if $l$ is a half-integer $l=\pm 1/2,\pm 3/2, ...\,$. 

In the case $k_z=0$, the Hamiltonian~\eqref{BdG} conserves a spin-orbital index, that is, $[H,\sigma_zs_z]=0$. We decompose the spinor basis in two spin-orbital blocks with $\sigma_zs_z \hat{\Psi}_{\pm}=\pm \hat{\Psi}_{\pm}$. Here $\hat{\Psi}_+=(\Psi_+,0)^T$, $\hat{\Psi}_-=(0,\Psi_-)^T$ where $\Psi_{+(-)}=\left(f_{1\uparrow(\downarrow)},f_{2\downarrow(\uparrow)},h_{1\downarrow(\uparrow)},h_{2\uparrow(\downarrow)}\right)^T$. After transformation given by Eq.~\eqref{major1}, we obtain
\begin{eqnarray}\label{H_trans}
\!\!\!\!\!\!H_{\rho}\!=\!\left(\!c_z m\!-\!\mu\!-\!\frac{lv}r c_x\!-\!
\rho iv  c_y \nabla_r\!\right)\!\tau_z\!-\!\Delta c_y \tau_x (ic_z)^{\nu}.
\end{eqnarray}
Here $\rho=\pm 1$ corresponds to different spin-orbital blocks and Pauli matrices $c_i$ act in the spin-orbital space $(1\uparrow,2\downarrow)$ for $\rho=+1$ and $(1\downarrow,2\uparrow)$ for $\rho=-1$, $\tau_i$ acts in the particle-hole space. 

The Hamiltonian~\eqref{BdG} has time-reversal, $T=is_yK$, and particle-hole conjugation, $\Xi=\sigma_y\tau_y K$, symmetries that combine into a chiral symmetry $U_c=\Xi T=i \tau_y$. The latter symmetry anti-commutes with the Hamiltonian, $\{H,U_c\}=0$. In the basis where the chiral operator $U_c$ is diagonal, the Hamiltonian transforms to
\begin{eqnarray}\label{H_t_pm}
H_t\!&=&\!
\begin{pmatrix}
0 & H_- \\
H_ + & 0
\end{pmatrix},\\
\nonumber
H_{\mp}\!&=&\!\!\mu\!+\! m \kappa_{\mp z}\!-\! \kappa_{\mp x}\frac{lv}r\!+\!i\left[\rho v\nabla_r\mp \Delta (i\kappa_{\mp z})^{\nu}\right]\!\kappa_{\mp y}.
\end{eqnarray} 
where $\kappa_{\mp i}$ are the Pauli matrices that act in the basis $\vec{L}=(L_1,L_2)^T=(h_{2\uparrow(\downarrow)}\mp i f_{1\downarrow(\uparrow)}, h_{1\downarrow(\uparrow)}\mp i f_{1\uparrow(\downarrow)})^T$ for $\rho=+1(-1)$. As a result, we decompose $8\times8$ system in four blocks of $2\times2$ equations. We can see, that $H_+$ differs from $H_-$ only by the sign before $\Delta$.

\section{Zero-energy solutions}\label{ZES}

First, we solve equations $H_{\mp}\vec{L}=0$ for the vortex of type I ($\nu=0$). The order parameter is eliminated by substitution $L_{1,2}=l_{1,2}\times\exp(\mp \rho\int dr\Delta/v)$. We get
\begin{eqnarray}\label{nu02}
\nonumber
v^2l_1''+\left[\mu^2-m^2-\frac{v^2l(l-\rho)}{r^2}\right]l_1=0,\\
(\mu-m)l_2 =v\!\left(\rho l_1'-\frac{l}{r}l_1\right),
\end{eqnarray}
 where the prime means the differentiation over $r$. Regular at $r=0$ solutions are
\begin{eqnarray}\label{vec}
\nonumber 
\vec{L}&=&Ne^{\mp\rho \int dr\Delta/v}\sqrt{r}\\
&\times&\begin{pmatrix}
\sqrt{\mu-m} J_{ l+\rho/2}(r\sqrt{\mu^2-m^2}/v)\\
\sqrt{\mu+m}J_{l-\rho/2}(r\sqrt{\mu^2-m^2}/v)
\end{pmatrix},
\end{eqnarray}
where $J_{\alpha}(x)$ are the Bessel functions and $N$ is a constant. We take into account that the strain and, consequently, the order parameter can be coordinate dependent. As we can see, $H_+$ has a normalized solution if $\rho=+1$ and $H_-$ has such a solution when $\rho=-1$.

The solutions with different signs of $\rho$ and $l$ are degenerate and form Kramer's pairs.
In the considered basis
\begin{eqnarray*}
\psi_1&=& [L_1(l),L_2(l),0,0,0,0,0,0],\\
\psi_2&=&(-1)^{2l} [0,0,0,0,0,0,L_1(-l),L_2(-l)]   
\end{eqnarray*}
are the components of such a pair, where $L_i$ is $i$-th component of the vector given by Eq.~\eqref{vec}. We can rewrite the obtained solutions in the original basis
\begin{eqnarray*}
\Psi_1&=&=[-iL_2(l),0,0,-iL_1(l),L_2(l),0,0,L_1(l)],\\
\Psi_2&=&=[0,iL_2(-l),iL_1(-l),0,0,L_2(-l),L_1(-l),0].   
\end{eqnarray*}
Since $J_n=(-1)^n J_{-n}$ for integer $n$ and $l$ is a half-integer, we obtain $\Psi_2=is_yK \Psi_1$. Thus, $\Psi_1$ and $\Psi_2$ are the Kramer's pair.

We can derive a dispersion of the obtained solution in $k_z$ using first order perturbation theory in $v_zk_z\sigma_y\tau_z$. For this goal, we have to calculate elements of the $4\times4$ matrix $N=\vec{\Psi_{i}} v_zk_z \sigma_y \tau_z \vec{\Psi_{j}}$. Note that the solutions with the same $\rho$ but different sign of the angular momenta, $\Psi_1(l)$ and $\Psi_1(-l)$, have different density of states. Nevertheless, a major of states is located at the distance $L \sim l\xi$ from the center of the vortex in both cases and in the limit $\mu \ll m$ the density of states for $\Psi_1(l)$ and $\Psi_1(-l)$ coincides. So, it is reasonable to consider only the matrix elements for the states with the same absolute value of the angular momenta $|l|$. If $\mu \ll m$, we find that the eigenvalues of $N$ consist of two doubly degenerate branches with $E=\pm v_zk_z$. Thus, the states in the spin vortex of types I have a linear dispersion in $z$ direction.

Similarly we consider the vortex of type II. We assume that $\Delta(r)=\Delta$ and for $\vec{L}=(L_1,L_2)$ we have
 \begin{eqnarray}\label{n11}
\nonumber
v^2L_1''&+&\left[\Delta^2+\mu^2-m^2\pm2i\Delta\frac{vl}{r}-\frac{v^2l(l-\rho)}{r^2}\right]L_1=0,\\
&&(\mu-m)L_2=v\left(\rho L_1'+\frac{l}{r}L_1\right)\pm i\Delta L_1.
\end{eqnarray}
The solution of this system can be expressed through the Whittaker's functions $M_{\beta,\gamma}(z)$ as $L_1=M_{\beta,\gamma}(2ir\sqrt{\mu^2+\Delta^2-m^2}/v)$ with $\beta=\pm l\Delta/\sqrt{\mu^2+\Delta^2-m^2}$ and $\gamma^2=1/4-l(l-\rho)$. The obtained solutions are regular at $r=0$, but are not regularized at $r \rightarrow +\infty$ since $L_1 \propto r^{|\beta|}\exp{\left[ir\sqrt{\mu^2+\Delta^2-m^2}/v\right]}$ at large radius. Thus, we conclude that no localized zero energy solutions exist in the vortex core of II type.  

The considered system belongs to the DIII symmetry class due to the presence of both time-reversal and particle-hole conjugation symmetries. This class is characterized by the topological invariant $Z_2$, which is associated with the time-reversal symmetry~\cite{Schnyder2008,Chiu2016}
\begin{equation}\label{Z2_gen}
Z_2=\prod\limits_{\bf K} \textrm{Pf}\left[\,w({\bf K})\right]/\sqrt{\det w({\bf K})}.
\end{equation}
Here Pf is a pfaffian, elements of a skew-symmetric matrix $w_{ij}(k)= \langle u_i(k) |\hat{T}| u_j(k) \rangle$ are calculated at time-reversal invariant momenta ${\bf K}$ in the reduced Brillouin zone, $u_i$ are eigenvectors of the Hamiltonian~\eqref{BdG} at $k_z=0$. We found (see for details Appendix C) that $Z_2$ is trivial for the vortex of type I and non-trivial for the vortex of type II
\begin{eqnarray}
Z_2=1\quad \textrm{for}\,\,\, \nu=0, \, \textrm{type I}\\
\nonumber
Z_2=-1\quad \textrm{for}\,\,\, \nu=1, \, \textrm{type II}.
\end{eqnarray}
Thus, the spin vortices of different types are topologically different.

\section{Discussion}

We obtain that two types of topologically different spin vortices can be induced by strain in the topological superconductors. The localized Majorana solutions of Hamiltonian~\eqref{BdG} exist near the core of the vortex of type I, while in the case of the vortex of type II such solutions does not observe. In Ref.~\cite{Wu2017} a similar result has been obtained for a different Hamiltonian with k-dependent nematic order parameter. We argue that this similarity arises due to the similarity of the topology of the spin vortices. That is, properties of the spin vortices are similar in different materials and are governed by $Z_2$ topological index. The spin vortex can be created by the application of a mechanical force applied at the center of a circular film of the doped topological insulator. The lattice strain caused by defects or substrate can also generate spin vortices. Since spin vortices have a normal core they can be detected by scanning tunneling microscopy or magnetic force microscopy.

Under the assumptions used above, an arbitrary small strain generates the spin vortex. This is a result of the degeneracy of the free energy of the nematic superconductor with respect to the nematicity direction $\alpha$. The degeneracy of the nematicity is commonly lifted by the presence of a strain or hexagonal warping~\cite{Fu2014}. The initial strain $u_0$, either spontaneous or arising during the crystal growth, is usually observed in the samples of the doped topological insulators~\cite{Akzyanov2020,Kuntsevich2019}. Thus, a large enough force should be applied to generate the spin vortex in a real experiment. In particular, the strain, $u$, produced by the applied force must be much larger than the initial strain $u_0$. Hexagonal warping fixes nematicity direction as well~\cite{Fu2014} and, consequently, prevents generation of the spin vortex. However, the corresponding terms appears in the GL free energy in the sixth order in the order parameter and are less relevant for fixing the nematicity than the strain. In principal, if the symmetry breaking field is smaller than spontaneous deformation~\cite{Akzyanov2020}, then, the nematicity direction becomes degenerate and the spin vortices can nucleate spontaneously. However, preparation of such samples with unfixed nematicity direction has not been reported so far.

The considered spin vortices have the normal core and, consequently, usual Caroli-de Gennes-Matricon states with non-zero energies $E_n$ exist near their centers. The spectrum of such quasiparticles for the doped topological insulator was calculated in Ref.~\onlinecite{Deng2020} in the quasiclassical approximation, $E_n=n \Delta^2/\sqrt{\mu^2-m^2}$, where $n=1,2,...$\,. However, the discussed here Majorana fermions with zero energy are a special type of the BdG solution and they do not require a normal vortex core to be localized. This can be seen from Eq.~\eqref{vec} (or from Refs.~\onlinecite{Akzyanov2015,Akzyanov2016a} for the case of emergent chiral symmetry). In particular, it is evidence that a particular form of the order parameter spatial dependence near the core of the vortex is not of importance for the existence of the Majorana solutions, and the assumption that $\Delta$ in BdG equations is independent of $r$ is a good approximation for zero-energy solutions.

In conclusion, we show that the rotational symmetric strain can generate spin vortices in the doped topological insulators. These vortices can be either of type I or II and have normal cores. The spin vortex of type I enhances superconductivity far from its core and generates localized zero-energy Majorana states, while the spin vortex of type II suppresses superconductivity and has no zero-energy states near its core. The different types of spin vortices have different topology. We establish a non-trivial relation between the strain and Majorana states in doped topological insulators.

\section*{Acknowledgments}
R.S.A. acknowledges the support by the Russian Scientific Foundation under Grant No. 20-72-00030 and partial support from the Foundation for the Advancement of Theoretical Physics and Mathematics “BASIS.” A.L.R. is supported by the Russian Science Foundation Grant No. 17-12-01544. Ginzburg-Landau calculations were performed under support of Russian Scientific Foundation Grant No. 17-12-01544, Bogoliubov-de Gennes equations were solved under the support of Russian Scientific Foundation Grant No. 20-72-00030.

\section*{Appendix A}\label{Appendix A}

Here we take into account the gradient terms in the GL functional. The main effect which is produced by these terms is the appearance of the normal core at the center of the vortex. Size of this vortex core is different for different types of the spin vortices. 

The GL free energy of the topological superconductor is a sum~\cite{Venderbos2016}
\begin{equation}\label{GL}
F=F_0+F_u+F_D,    
\end{equation}
where $F_0$ is the GL free energy in the absence of the strain, \eqref{F0}, and $F_u$ is the contribution due to strain, \eqref{Fu}. The term $F_D$ arises due to inhomogeneity of the order parameter. We consider here the case of nematic order, that is, the $B_2>0$ in $F_0$ and the GL order parameter is real, $\vec{\Delta}=\Delta(\cos \alpha, \sin \alpha)$. In this case Eq.~(3) from Ref.~\cite{Venderbos2016} can be rewritten as
\begin{widetext}
\begin{eqnarray*}\label{GL_D}
F_D &=& -(J_1+J_4)\left[\left(\frac{\partial\Delta_1}{\partial x}\right)^2+\left(\frac{\partial\Delta_2}{\partial y}\right)^2\right]
-(J_1-J_4)\left[\left(\frac{\partial\Delta_1}{\partial y}\right)^2-\left(\frac{\partial\Delta_2}{\partial x}\right)^2\right]\\
&-&2(J_4+J_2)\frac{\partial\Delta_1}{\partial x}\frac{\partial\Delta_2}{\partial y}-2(J_4-J_2)\frac{\partial\Delta_1}{\partial y}\frac{\partial\Delta_2}{\partial x}
-J_3\left[\left(\frac{\partial\Delta_1}{\partial z}\right)^2+\left(\frac{\partial\Delta_2}{\partial z}\right)^2\right].
\end{eqnarray*}
Corresponding GL equations are
\begin{eqnarray}\label{GL12}
\delta F_i&=&\frac{\delta F}{\delta \Delta_i^*}=0,\qquad i=1,2,\\
\nonumber
\delta F_1&=&A\Delta_1\!+\!2B_1(\Delta_1^2\!+\!\eta_2^2)\Delta_1\!+\!g_N\left[(u_{xx}
\!-\!u_{yy})\Delta_1\!+\!2u_{xy}\Delta_{2}\right]\!+\!
J_1(\partial_x^2\!+\!\partial_y^2)\Delta_1\!+\!J_3\partial_z^2\Delta_1\!+\!J_4\left[(\partial_x^2\!-\!\partial_y^2)\Delta_1\!+\!2\partial_x\partial_y\Delta_{2}\right],\\ 
\nonumber
\delta F_2&=&A\Delta_2\!+\!2B_1(\Delta_1^2\!+\!\eta_2^2)\Delta_2\!+\!g_N\left[-(u_{xx}\!-\!u_{yy})\Delta_{2}\!+\!2u_{xy}\eta_1\right]\! +\!
J_1(\partial_x^2\!+\!\partial_y^2)\Delta_2\!+\!J_3\partial_z^2\Delta_2\!+\!J_4\left[-(\partial_x^2-\partial_y^2)\Delta_2\!+\!2\partial_x\partial_y\Delta_{1}\right]. \end{eqnarray}
\end{widetext}

Let a central symmetric force along $z$ direction acts on a plate of the topological insulator, which has a form of disc. We introduce cylindrical coordinates $(x,y,z)=(r\cos \varphi, r\sin \varphi,z)$. The force $\vec{f}=\left[0,0,f(r,z)\right]$ produces an elastic displacement of the material with components $u_r(r,z)$ and $u_z(r,z)$. An elementary algebra allows us to express components of the strain tensor $u_{ij}$ in terms of $u_r(r,z)$ and $u_z(r,z)$ and their derivatives 
\begin{eqnarray}\label{strain1}
u_{xx}&-&u_{yy}=u(r,z)\cos 2\varphi, \\
\nonumber
2u_{xy}&=&u(r,z)\sin 2\varphi,\,\,\,
u(r,z)=\frac{\partial u_r}{\partial r}+\frac{u_r}{r}.
\end{eqnarray}
The value $u(r,z)$ depends on $f(r,z)$ and on the boundary conditions of a particular elastic problem. However, $u_r(0,z)=0$ in any case due to the problem central symmetry.

We assume that the angular symmetries of the vortex near and far from the core [see Eqs.~(8) and (9)] are similar. Thus, we will seek solutions of the GL equations~\eqref{GL12} in the form
\begin{eqnarray}\label{nematic}
\vec{\Delta}&=&\Delta(r,z)\left[\cos\alpha(\varphi),\sin\alpha(\varphi)\right],\\
\nonumber
 &&\alpha(\varphi)=m\varphi+\phi_0.
\end{eqnarray}
Here $m$ and $\phi_0$ are real and $\Delta(r,z)$ is positive or zero. 

We introduce linear combinations
\begin{eqnarray}\label{GLcomb}
\nonumber
\delta F_{\alpha}=\delta F_1\sin \alpha\!-\!\delta F_2\cos \alpha,\,\, 
\delta F_\Delta=\delta F_1\cos \alpha\!+\!\delta F_2\sin \alpha.
\end{eqnarray}
In the cylindrical coordinates equation for $\delta F_{\alpha}$ is
\begin{widetext}
\begin{eqnarray}\label{GLalpha}
\delta F_{\alpha}=\sin{\left[2\phi_0+2(m-1)\varphi\right]}\left\{J_4\left[\Delta''(r)-\frac{2m-1}{r}\Delta'(r)
+\frac{m(m+2)}{r^2}\Delta(r)\right]+g_Nu\Delta(r)\right\}=0, 
\end{eqnarray}
\end{widetext}
where prime means the differentiation by $r$. This equation is compatible with $\delta F_\Delta=0$ only if $m=1$ and $\phi_0=0$ (the spin vortex of type I) or $\phi_0=\pi/2$ (the spin vortex of type II). The second GL equation then reads  
\begin{widetext}
\begin{eqnarray}\label{GLeta}
\delta F_\Delta = \left(J_1\pm J_4\right)\left(\Delta''+\frac{1}{r}\Delta'-\frac{1}{r^2}\Delta\right)
+J_3\frac{\partial^2\Delta}{\partial z^2}+2B_1\Delta^3+\left(A \pm g_Nu\right)\Delta=0.
\end{eqnarray}
\end{widetext}
The spatial scale of variation of $\Delta$ in the $z$-direction is dictated by (macroscopic) elastic part of the problem, and it is much larger than the scale in the $r$-direction near the center (core) of the vortex which is of the order of (microscopic) coherence length of superconductivity. Therefore, the value $\partial^2\Delta/\partial z^2$ is small and we ignore $z$-dependence of the order parameter. Under the latter assumption we rewrite Eq.~\eqref{GLeta} in dimensionless variables as   
\begin{eqnarray}\label{GLeta_dim}
\nonumber
f''(\bar{r})&+&\frac{1}{\bar{r}}f'(\bar{r})-\left[1+\frac{1}{\bar{r^2}}f(\bar{r})\right]+f(\bar{r})^3=0, \\
\nonumber
f(r)&=&\frac{\Delta(r)}{\Delta_0},\quad \Delta_0=\sqrt{\frac{\pm g_Nu-A}{2B_1}},\\
\bar{r}&=&r/\xi_{I,II}, \quad \xi_{I,II}=\sqrt{\frac{\pm g_Nu-A}{J_1 \pm J_4}}.
\end{eqnarray}
Here $\xi_{I}$ and $\xi_{II}$ are effective coherence lengths for the vortex of type I and for the vortex of type II, respectively. These values differ from the coherence $\xi_0$ in the sample without strain. 

We neglect the coordinate dependence of $\Delta_0$ since its scale is of the order of the spatial scale of the elastic strain and is much larger than $\xi_i$. The latter equation is the same as the equation for the order parameter in an `ordinary' superconductor near core of the Abrikosov vortex~\cite{Abrikosov1957}. Thus, the behaviour of the order parameter in the case of the spin vortex is similar to that in the case of the Abrikosov vortex. The order parameter is zero at $r=0$, increases linearly in the region $\bar{r} \ll 1$ and is equal to $\Delta_0$ when $\bar{r} \gg 1$. Thus, the spin vortex has a normal core which can be considered as a topological defect. The size of this normal core is different for different types of the spin vortices.

\section*{Appendix B}

Here we give a derivation of the equations used in Sections~\ref{Section_BdG} and~\ref{ZES} with more technical details. We start with the BdG Hamiltonian~\eqref{BdG} rewritten for convenience in the form
\begin{eqnarray*}
\nonumber
H_{\textrm{BdG}}(\mathbf{k})&=&-\mu+ m\sigma_z+v\sigma_x(s_xk_y-s_yk_x)\\
&+&v_zk_zs_x\sigma_y\tau_z+(\Delta_x s_x+\Delta_ys_y)\sigma_y\tau_x,
\end{eqnarray*}
We put first $k_z=0$. In the polar coordinate space components of the momentum operator are
    \begin{eqnarray*}
    k_x&=&-i(\nabla_r \cos \varphi - \sin \varphi/r \,\nabla_{\varphi}), \\
     k_y&=&-i(\nabla_r  \sin \varphi + \cos \varphi/r \,\nabla_{\varphi}).
    \end{eqnarray*}
We substitute these operators to Hamiltonian~\eqref{BdG} and come to Eq.~\eqref{Bdg1}. As it was mentioned in Section~\ref{Section_BdG}, this Hamiltonian conserves spin-orbital index, $[H,s_z\sigma_z]=0$. In this case, there exists a basis, in which eigenvectors $u_i^\pm$  of the Hamiltonian are classified according to this index, that is, $Hu_i^\pm=\varepsilon_i u_i^\pm$ and $s_z \sigma_z u_i^+ = + u_i$ and  $s_z \sigma_z u_i^- = - u_i$. The Hamiltonian is block diagonal in a basis where the operator of a conserved index is diagonal. Thus, we can choose the basis where each block of the Hamiltonian corresponds to the eigenvectors with the same index (plus or minus) of the spin-orbit operator. The spin orbital operator $s_z\sigma_z$ is already diagonal and we need simply to rearrange strings of the Hamiltonian to make it block-diagonal. It can be done by transformation 
\begin{equation}\label{W}
W=\left(
\begin{array}{cccccccc}
 1 & 0 & 0 & 0 & 0 & 0 & 0 & 0 \\
 0 & 0 & 0 & 0 & 1 & 0 & 0 & 0 \\
 0 & 0 & 0 & 0 & 0 & 1 & 0 & 0 \\
 0 & 1 & 0 & 0 & 0 & 0 & 0 & 0 \\
 0 & 0 & 1 & 0 & 0 & 0 & 0 & 0 \\
 0 & 0 & 0 & 0 & 0 & 0 & 1 & 0 \\
 0 & 0 & 0 & 0 & 0 & 0 & 0 & 1 \\
 0 & 0 & 0 & 1 & 0 & 0 & 0 & 0 \\
\end{array}
\right).
\end{equation}
We apply this transformation to the Hamiltonian, $W^{\dagger}H(r,l)W$, and get
\begin{equation}\label{hfirst}
H=\left(
\begin{array}{cc}
H_+ &0 \\ 
0  &H_-
\end{array}\right)
\end{equation}
where $0$ in the latter expression corresponds to $4\times 4$ zero matrix and $H_\rho=H_\pm$ is determined by Eq.~\eqref{H_trans}.

The Hamiltonian $H_{\rho}$ anticommutes with $U_c=i\tau_y$, $\{H_{\rho},U_c\}=0$. In the basis where $U_c$ is diagonal, the Hamiltonian $H_{\rho}$ will be off block diagonal matrix. The operator $i\tau_y$ is diagonalized by the transformation
\begin{equation}
R=\frac{1}{\sqrt{2}}\left(
\begin{array}{cccc}
 0 & -i & 0 & i \\
 -i & 0 & i & 0 \\
 0 & 1 & 0 & 1 \\
 1 & 0 & 1 & 0 \\
\end{array}
\right).
\end{equation}
We apply this transformation to the Hamiltonian, $H_t=R^{\dagger}H_{\rho}R$, and obtain Eqs.~\eqref{H_t_pm}.

\section{Appendix C}

Nematic superconductor with the spin vortex belongs to the DIII symmetry class that is classified by $Z_2$ topological invariant Eq.~\eqref{Z2_gen}. Here we calculate this value following Ref.~\onlinecite{Chiu2016}. In formula~\eqref{Z2_gen} the elements of a skew-symmetric matrix $w_{ij}(k)= \langle u_i(k)| \hat{T}| u_j(k) \rangle$ are calculated at time-reversal invariant momenta ${\bf K}$ in the reduced Brillouin zone, $u_i$ are eigenvectors of the Hamiltonian given by Eq.~\eqref{BdG} for $k_z=0$, and $\hat{T}=is_yK$ is the operator of the time-reversal symmetry. We use here the basis in which the  Hamiltonian has the form~\eqref{hfirst}. The operator of the time-reversal symmetry in this basis is $\tilde{T}=W^{T}s_yW iK=ic_zt_yK$, where $W$ is given by Eq.~\eqref{W}. Explicitly,
\begin{equation}\label{T}
\tilde{T}=\left(
\begin{array}{cccccccc}
 0 & 0 & 0 & 0 &-i & 0 & 0 & 0 \\
 0 & 0 & 0 & 0 & 0 & i & 0 & 0 \\
 0 & 0 & 0 & 0 & 0 & 0 &-i & 0 \\
 0 & 0 & 0 & 0 & 0 & 0 & 0 & i \\
 i & 0 & 0 & 0 & 0 & 0 & 0 & 0 \\
 0 & -i& 0 & 0 & 0 & 0 & 0 & 0 \\
 0 & 0 & i& 0 & 0 & 0 & 0 & 0 \\
 0 & 0 & 0 &-i & 0 & 0 & 0 & 0 \\
\end{array}
\right)iK.
\end{equation}
In these terms, the Hamiltonian can be  rewritten as
\begin{eqnarray}\label{H_trans3}
H(k)&=&
\begin{pmatrix}
H_{\rho=+1} &0 \\
0 & H_{\rho=-1}
\end{pmatrix},\\
\nonumber
H_{\rho}&=&\left(\!c_z m\!-\!\mu\!-\rho vk_xc_y+vk_yc_x\!\right)\!\tau_z\\
\nonumber
&+&\Delta c_y \tau_x\cos{\varphi}-\rho \Delta c_x \tau_x\sin{\varphi}.
\end{eqnarray}
Eigenvectors of the Hamiltonian are $P_{+(-)}=(\psi_+,0)$ and $(0,\psi_-)$ where $\psi_{+(-)}$ are the eigenvectors of $H_{\rho=+1(-1)}$. Matrix elements $P_i \tilde{T} P_j$ for the eigenvectors with the same $\rho$ vanishes and only the states with different $\rho$ contribute to $w({\bf K})$, that is, $\langle P_+ \tilde{T} P_-\rangle=\langle \psi_{+i}c_z\psi_{-j}\rangle$ and $\langle P_- \tilde{T} P_+\rangle=-\langle \psi_{-i}c_z\psi_{+j}\rangle$. The skew-symmetric matrix $w_{ij}$ reads
\begin{eqnarray}
w(k_x,k_y)=
\begin{pmatrix}
0 &Q(k_x,k_y) \\
-Q^T(k_x,k_y) & 0
\end{pmatrix},
\end{eqnarray}
where $Q_{ij}(k)=\langle \psi_{+i}(k)| c_z K| \psi_{-j} \rangle$, $i,j=1,..,4$. Using a well-known formula for the Pffafian, we get
\begin{equation}
Z_2=\prod\limits_{\bf K}\textrm{Det}\,Q({\bf k}).
\end{equation}
This product is calculated at the time-reversal momenta ${\bf K}$ of the reduced Brillouin zone.
Explicit calculation for spin vortex of the type I gives that $\textrm{Det}\,Q(k_x,k_y)=1$. Thus, the topological index is trivial in this case, $Z_2=+1$. For spin vortex of the type II we get $\textrm{Det}\,Q(k_x,k_y)=q(k_x,k_y)\textrm{sign}\,(k_x^2v^2+k_y^2v^2+m^2-\mu^2-\Delta^2)$. The sign of the latter value is different for small and large momenta. We have to calculate the determinant at the time-reversal symmetric points of the reduced Brillouin zone. Thus, we have 
$Z_2=q(0,0)q(+\infty,0)q(0,+\infty),q(+\infty,+\infty)=-1$. 
We see that the spin vortex of type II has a non-trivial topology, which is different from the topology of the spin vortex of type I.

\bibliography{strain}

\begin{thebibliography}{34}%
\makeatletter
\providecommand \@ifxundefined [1]{%
 \@ifx{#1\undefined}
}%
\providecommand \@ifnum [1]{%
 \ifnum #1\expandafter \@firstoftwo
 \else \expandafter \@secondoftwo
 \fi
}%
\providecommand \@ifx [1]{%
 \ifx #1\expandafter \@firstoftwo
 \else \expandafter \@secondoftwo
 \fi
}%
\providecommand \natexlab [1]{#1}%
\providecommand \enquote  [1]{``#1''}%
\providecommand \bibnamefont  [1]{#1}%
\providecommand \bibfnamefont [1]{#1}%
\providecommand \citenamefont [1]{#1}%
\providecommand \href@noop [0]{\@secondoftwo}%
\providecommand \href [0]{\begingroup \@sanitize@url \@href}%
\providecommand \@href[1]{\@@startlink{#1}\@@href}%
\providecommand \@@href[1]{\endgroup#1\@@endlink}%
\providecommand \@sanitize@url [0]{\catcode `\\12\catcode `\$12\catcode
  `\&12\catcode `\#12\catcode `\^12\catcode `\_12\catcode `\%12\relax}%
\providecommand \@@startlink[1]{}%
\providecommand \@@endlink[0]{}%
\providecommand \url  [0]{\begingroup\@sanitize@url \@url }%
\providecommand \@url [1]{\endgroup\@href {#1}{\urlprefix }}%
\providecommand \urlprefix  [0]{URL }%
\providecommand \Eprint [0]{\href }%
\providecommand \doibase [0]{https://doi.org/}%
\providecommand \selectlanguage [0]{\@gobble}%
\providecommand \bibinfo  [0]{\@secondoftwo}%
\providecommand \bibfield  [0]{\@secondoftwo}%
\providecommand \translation [1]{[#1]}%
\providecommand \BibitemOpen [0]{}%
\providecommand \bibitemStop [0]{}%
\providecommand \bibitemNoStop [0]{.\EOS\space}%
\providecommand \EOS [0]{\spacefactor3000\relax}%
\providecommand \BibitemShut  [1]{\csname bibitem#1\endcsname}%
\let\auto@bib@innerbib\@empty
%</preamble>
\bibitem [{\citenamefont {Yonezawa}(2018)}]{Yonezawa2018}%
  \BibitemOpen
  \bibfield  {author} {\bibinfo {author} {\bibfnamefont {S.}~\bibnamefont
  {Yonezawa}},\ }\bibfield  {title} {\bibinfo {title} {Nematic
  superconductivity in doped bi2se3 topological superconductors},\ }\href
  {https://doi.org/10.3390/condmat4010002} {\bibfield  {journal} {\bibinfo
  {journal} {Condensed Matter}\ }\textbf {\bibinfo {volume} {4}},\ \bibinfo
  {pages} {2} (\bibinfo {year} {2018})}\BibitemShut {NoStop}%
\bibitem [{\citenamefont {Cho}\ \emph {et~al.}(2020)\citenamefont {Cho},
  \citenamefont {Shen}, \citenamefont {Lyu}, \citenamefont {Atanov},
  \citenamefont {Chen}, \citenamefont {Lee}, \citenamefont {Hor}, \citenamefont
  {Gawryluk}, \citenamefont {Pomjakushina}, \citenamefont {Bartkowiak},
  \citenamefont {Hecker}, \citenamefont {Schmalian},\ and\ \citenamefont
  {Lortz}}]{Cho2020}%
  \BibitemOpen
  \bibfield  {author} {\bibinfo {author} {\bibfnamefont {C.-w.}\ \bibnamefont
  {Cho}}, \bibinfo {author} {\bibfnamefont {J.}~\bibnamefont {Shen}}, \bibinfo
  {author} {\bibfnamefont {J.}~\bibnamefont {Lyu}}, \bibinfo {author}
  {\bibfnamefont {O.}~\bibnamefont {Atanov}}, \bibinfo {author} {\bibfnamefont
  {Q.}~\bibnamefont {Chen}}, \bibinfo {author} {\bibfnamefont {S.~H.}\
  \bibnamefont {Lee}}, \bibinfo {author} {\bibfnamefont {Y.~S.}\ \bibnamefont
  {Hor}}, \bibinfo {author} {\bibfnamefont {D.~J.}\ \bibnamefont {Gawryluk}},
  \bibinfo {author} {\bibfnamefont {E.}~\bibnamefont {Pomjakushina}}, \bibinfo
  {author} {\bibfnamefont {M.}~\bibnamefont {Bartkowiak}}, \bibinfo {author}
  {\bibfnamefont {M.}~\bibnamefont {Hecker}}, \bibinfo {author} {\bibfnamefont
  {J.}~\bibnamefont {Schmalian}},\ and\ \bibinfo {author} {\bibfnamefont
  {R.}~\bibnamefont {Lortz}},\ }\bibfield  {title} {\bibinfo {title}
  {Z3-vestigial nematic order due to superconducting fluctuations in the doped
  topological insulators nbxbi2se3 and cuxbi2se3},\ }\href
  {https://doi.org/10.1038/s41467-020-16871-9} {\bibfield  {journal} {\bibinfo
  {journal} {Nature Communications}\ }\textbf {\bibinfo {volume} {11}},\
  \bibinfo {pages} {3056} (\bibinfo {year} {2020})}\BibitemShut {NoStop}%
\bibitem [{\citenamefont {Fr\"ohlich}\ \emph {et~al.}(2020)\citenamefont
  {Fr\"ohlich}, \citenamefont {Wang}, \citenamefont {Bagchi}, \citenamefont
  {Stunault}, \citenamefont {Ando},\ and\ \citenamefont
  {Braden}}]{Frohlich2020}%
  \BibitemOpen
  \bibfield  {author} {\bibinfo {author} {\bibfnamefont {T.}~\bibnamefont
  {Fr\"ohlich}}, \bibinfo {author} {\bibfnamefont {Z.}~\bibnamefont {Wang}},
  \bibinfo {author} {\bibfnamefont {M.}~\bibnamefont {Bagchi}}, \bibinfo
  {author} {\bibfnamefont {A.}~\bibnamefont {Stunault}}, \bibinfo {author}
  {\bibfnamefont {Y.}~\bibnamefont {Ando}},\ and\ \bibinfo {author}
  {\bibfnamefont {M.}~\bibnamefont {Braden}},\ }\bibfield  {title} {\bibinfo
  {title} {Crystal structure and distortion of superconducting
  ${\mathrm{cu}}_{x}{\mathrm{bi}}_{2}{\mathrm{se}}_{3}$},\ }\href
  {https://doi.org/10.1103/PhysRevMaterials.4.054802} {\bibfield  {journal}
  {\bibinfo  {journal} {Phys. Rev. Materials}\ }\textbf {\bibinfo {volume}
  {4}},\ \bibinfo {pages} {054802} (\bibinfo {year} {2020})}\BibitemShut
  {NoStop}%
\bibitem [{\citenamefont {Schmidt}\ \emph {et~al.}(2020)\citenamefont
  {Schmidt}, \citenamefont {Parhizgar},\ and\ \citenamefont
  {Black-Schaffer}}]{Schmidt2020}%
  \BibitemOpen
  \bibfield  {author} {\bibinfo {author} {\bibfnamefont {J.}~\bibnamefont
  {Schmidt}}, \bibinfo {author} {\bibfnamefont {F.}~\bibnamefont {Parhizgar}},\
  and\ \bibinfo {author} {\bibfnamefont {A.~M.}\ \bibnamefont
  {Black-Schaffer}},\ }\bibfield  {title} {\bibinfo {title} {Odd-frequency
  superconductivity and meissner effect in the doped topological insulator
  ${\mathrm{bi}}_{2}{\mathrm{se}}_{3}$},\ }\href
  {https://doi.org/10.1103/PhysRevB.101.180512} {\bibfield  {journal} {\bibinfo
   {journal} {Phys. Rev. B}\ }\textbf {\bibinfo {volume} {101}},\ \bibinfo
  {pages} {180512(R)} (\bibinfo {year} {2020})}\BibitemShut {NoStop}%
\bibitem [{\citenamefont {Das}\ \emph {et~al.}(2020)\citenamefont {Das},
  \citenamefont {Kobayashi}, \citenamefont {Smylie}, \citenamefont {Mielke},
  \citenamefont {Takahashi}, \citenamefont {Willa}, \citenamefont {Yin},
  \citenamefont {Welp}, \citenamefont {Hasan}, \citenamefont {Amato},
  \citenamefont {Luetkens},\ and\ \citenamefont {Guguchia}}]{Das2020}%
  \BibitemOpen
  \bibfield  {author} {\bibinfo {author} {\bibfnamefont {D.}~\bibnamefont
  {Das}}, \bibinfo {author} {\bibfnamefont {K.}~\bibnamefont {Kobayashi}},
  \bibinfo {author} {\bibfnamefont {M.~P.}\ \bibnamefont {Smylie}}, \bibinfo
  {author} {\bibfnamefont {C.}~\bibnamefont {Mielke}}, \bibinfo {author}
  {\bibfnamefont {T.}~\bibnamefont {Takahashi}}, \bibinfo {author}
  {\bibfnamefont {K.}~\bibnamefont {Willa}}, \bibinfo {author} {\bibfnamefont
  {J.-X.}\ \bibnamefont {Yin}}, \bibinfo {author} {\bibfnamefont
  {U.}~\bibnamefont {Welp}}, \bibinfo {author} {\bibfnamefont {M.~Z.}\
  \bibnamefont {Hasan}}, \bibinfo {author} {\bibfnamefont {A.}~\bibnamefont
  {Amato}}, \bibinfo {author} {\bibfnamefont {H.}~\bibnamefont {Luetkens}},\
  and\ \bibinfo {author} {\bibfnamefont {Z.}~\bibnamefont {Guguchia}},\
  }\bibfield  {title} {\bibinfo {title} {Time-reversal invariant and fully
  gapped unconventional superconducting state in the bulk of the topological
  compound ${\mathrm{nb}}_{0.25}{\mathrm{bi}}_{2}{\mathrm{se}}_{3}$},\ }\href
  {https://doi.org/10.1103/PhysRevB.102.134514} {\bibfield  {journal} {\bibinfo
   {journal} {Phys. Rev. B}\ }\textbf {\bibinfo {volume} {102}},\ \bibinfo
  {pages} {134514} (\bibinfo {year} {2020})}\BibitemShut {NoStop}%
\bibitem [{\citenamefont {Kuntsevich}\ \emph {et~al.}(2020)\citenamefont
  {Kuntsevich}, \citenamefont {Rybalchenko}, \citenamefont {Martovitskii},
  \citenamefont {Bannikov}, \citenamefont {Selivanov}, \citenamefont
  {Gavrilkin}, \citenamefont {Tsvetkov},\ and\ \citenamefont
  {Chizhevskii}}]{Kuntsevich2020}%
  \BibitemOpen
  \bibfield  {author} {\bibinfo {author} {\bibfnamefont {A.~Y.}\ \bibnamefont
  {Kuntsevich}}, \bibinfo {author} {\bibfnamefont {G.~V.}\ \bibnamefont
  {Rybalchenko}}, \bibinfo {author} {\bibfnamefont {V.~P.}\ \bibnamefont
  {Martovitskii}}, \bibinfo {author} {\bibfnamefont {M.~I.}\ \bibnamefont
  {Bannikov}}, \bibinfo {author} {\bibfnamefont {Y.~G.}\ \bibnamefont
  {Selivanov}}, \bibinfo {author} {\bibfnamefont {S.~Y.}\ \bibnamefont
  {Gavrilkin}}, \bibinfo {author} {\bibfnamefont {A.~Y.}\ \bibnamefont
  {Tsvetkov}},\ and\ \bibinfo {author} {\bibfnamefont {E.~G.}\ \bibnamefont
  {Chizhevskii}},\ }\bibfield  {title} {\bibinfo {title} {Effect of barium
  codoping on superconductivity in srxbi2se3},\ }\href
  {https://doi.org/10.1134/S002136402003008X} {\bibfield  {journal} {\bibinfo
  {journal} {JETP Letters}\ }\textbf {\bibinfo {volume} {111}},\ \bibinfo
  {pages} {151} (\bibinfo {year} {2020})}\BibitemShut {NoStop}%
\bibitem [{\citenamefont {Akzyanov}\ \emph
  {et~al.}(2020{\natexlab{a}})\citenamefont {Akzyanov}, \citenamefont
  {Kapranov},\ and\ \citenamefont {Rakhmanov}}]{Akzyanov2020}%
  \BibitemOpen
  \bibfield  {author} {\bibinfo {author} {\bibfnamefont {R.~S.}\ \bibnamefont
  {Akzyanov}}, \bibinfo {author} {\bibfnamefont {A.~V.}\ \bibnamefont
  {Kapranov}},\ and\ \bibinfo {author} {\bibfnamefont {A.~L.}\ \bibnamefont
  {Rakhmanov}},\ }\bibfield  {title} {\bibinfo {title} {Spontaneous strain and
  magnetization in doped topological insulators with nematic and chiral
  superconductivity},\ }\href {https://doi.org/10.1103/PhysRevB.102.100505}
  {\bibfield  {journal} {\bibinfo  {journal} {Phys. Rev. B}\ }\textbf {\bibinfo
  {volume} {102}},\ \bibinfo {pages} {100505(R)} (\bibinfo {year}
  {2020}{\natexlab{a}})}\BibitemShut {NoStop}%
\bibitem [{\citenamefont {Akzyanov}\ \emph
  {et~al.}(2020{\natexlab{b}})\citenamefont {Akzyanov}, \citenamefont
  {Khokhlov},\ and\ \citenamefont {Rakhmanov}}]{Akzyanov2020a}%
  \BibitemOpen
  \bibfield  {author} {\bibinfo {author} {\bibfnamefont {R.~S.}\ \bibnamefont
  {Akzyanov}}, \bibinfo {author} {\bibfnamefont {D.~A.}\ \bibnamefont
  {Khokhlov}},\ and\ \bibinfo {author} {\bibfnamefont {A.~L.}\ \bibnamefont
  {Rakhmanov}},\ }\bibfield  {title} {\bibinfo {title} {Nematic
  superconductivity in topological insulators induced by hexagonal warping},\
  }\href {https://doi.org/10.1103/PhysRevB.102.094511} {\bibfield  {journal}
  {\bibinfo  {journal} {Phys. Rev. B}\ }\textbf {\bibinfo {volume} {102}},\
  \bibinfo {pages} {094511} (\bibinfo {year} {2020}{\natexlab{b}})}\BibitemShut
  {NoStop}%
\bibitem [{\citenamefont {Fu}(2009)}]{Fu2009}%
  \BibitemOpen
  \bibfield  {author} {\bibinfo {author} {\bibfnamefont {L.}~\bibnamefont
  {Fu}},\ }\bibfield  {title} {\bibinfo {title} {Hexagonal warping effects in
  the surface states of the topological insulator
  ${\mathrm{bi}}_{2}{\mathrm{te}}_{3}$},\ }\href
  {https://doi.org/10.1103/PhysRevLett.103.266801} {\bibfield  {journal}
  {\bibinfo  {journal} {Phys. Rev. Lett.}\ }\textbf {\bibinfo {volume} {103}},\
  \bibinfo {pages} {266801} (\bibinfo {year} {2009})}\BibitemShut {NoStop}%
\bibitem [{\citenamefont {Fu}(2014)}]{Fu2014}%
  \BibitemOpen
  \bibfield  {author} {\bibinfo {author} {\bibfnamefont {L.}~\bibnamefont
  {Fu}},\ }\bibfield  {title} {\bibinfo {title} {Odd-parity topological
  superconductor with nematic order: Application to
  ${\mathrm{cu}}_{x}{\mathrm{bi}}_{2}{\mathrm{se}}_{3}$},\ }\href
  {https://link.aps.org/doi/10.1103/PhysRevB.90.100509} {\bibfield  {journal}
  {\bibinfo  {journal} {Phys. Rev. B}\ }\textbf {\bibinfo {volume} {90}},\
  \bibinfo {pages} {100509(R)} (\bibinfo {year} {2014})}\BibitemShut {NoStop}%
\bibitem [{\citenamefont {Matano}\ \emph {et~al.}(2016)\citenamefont {Matano},
  \citenamefont {Kriener}, \citenamefont {Segawa}, \citenamefont {Ando},\ and\
  \citenamefont {qing Zheng}}]{Matano2016}%
  \BibitemOpen
  \bibfield  {author} {\bibinfo {author} {\bibfnamefont {K.}~\bibnamefont
  {Matano}}, \bibinfo {author} {\bibfnamefont {M.}~\bibnamefont {Kriener}},
  \bibinfo {author} {\bibfnamefont {K.}~\bibnamefont {Segawa}}, \bibinfo
  {author} {\bibfnamefont {Y.}~\bibnamefont {Ando}},\ and\ \bibinfo {author}
  {\bibfnamefont {G.}~\bibnamefont {qing Zheng}},\ }\bibfield  {title}
  {\bibinfo {title} {Spin-rotation symmetry breaking in the superconducting
  state of {CuxBi}2se3},\ }\href {https://doi.org/10.1038/nphys3781} {\bibfield
   {journal} {\bibinfo  {journal} {Nature Physics}\ }\textbf {\bibinfo {volume}
  {12}},\ \bibinfo {pages} {852} (\bibinfo {year} {2016})}\BibitemShut
  {NoStop}%
\bibitem [{\citenamefont {Fu}\ and\ \citenamefont {Kane}(2008)}]{Fu2008}%
  \BibitemOpen
  \bibfield  {author} {\bibinfo {author} {\bibfnamefont {L.}~\bibnamefont
  {Fu}}\ and\ \bibinfo {author} {\bibfnamefont {C.~L.}\ \bibnamefont {Kane}},\
  }\bibfield  {title} {\bibinfo {title} {Superconducting proximity effect and
  majorana fermions at the surface of a topological insulator},\ }\href
  {https://link.aps.org/doi/10.1103/PhysRevLett.100.096407} {\bibfield
  {journal} {\bibinfo  {journal} {Phys. Rev. Lett.}\ }\textbf {\bibinfo
  {volume} {100}},\ \bibinfo {pages} {096407} (\bibinfo {year}
  {2008})}\BibitemShut {NoStop}%
\bibitem [{\citenamefont {Qi}\ and\ \citenamefont {Zhang}(2011)}]{Qi2011}%
  \BibitemOpen
  \bibfield  {author} {\bibinfo {author} {\bibfnamefont {X.-L.}\ \bibnamefont
  {Qi}}\ and\ \bibinfo {author} {\bibfnamefont {S.-C.}\ \bibnamefont {Zhang}},\
  }\bibfield  {title} {\bibinfo {title} {Topological insulators and
  superconductors},\ }\href {https://doi.org/10.1103/RevModPhys.83.1057}
  {\bibfield  {journal} {\bibinfo  {journal} {Rev. Mod. Phys.}\ }\textbf
  {\bibinfo {volume} {83}},\ \bibinfo {pages} {1057} (\bibinfo {year}
  {2011})}\BibitemShut {NoStop}%
\bibitem [{\citenamefont {Teo}\ and\ \citenamefont {Kane}(2010)}]{Teo2010}%
  \BibitemOpen
  \bibfield  {author} {\bibinfo {author} {\bibfnamefont {J.~C.~Y.}\
  \bibnamefont {Teo}}\ and\ \bibinfo {author} {\bibfnamefont {C.~L.}\
  \bibnamefont {Kane}},\ }\bibfield  {title} {\bibinfo {title} {Topological
  defects and gapless modes in insulators and superconductors},\ }\href
  {https://doi.org/10.1103/PhysRevB.82.115120} {\bibfield  {journal} {\bibinfo
  {journal} {Phys. Rev. B}\ }\textbf {\bibinfo {volume} {82}},\ \bibinfo
  {pages} {115120} (\bibinfo {year} {2010})}\BibitemShut {NoStop}%
\bibitem [{\citenamefont {Volovik}(1999)}]{Volovik1999}%
  \BibitemOpen
  \bibfield  {author} {\bibinfo {author} {\bibfnamefont {G.~E.}\ \bibnamefont
  {Volovik}},\ }\bibfield  {title} {\bibinfo {title} {Fermion zero modes on
  vortices in chiral superconductors},\ }\href
  {https://doi.org/doi.org/10.1134/1.568223} {\bibfield  {journal} {\bibinfo
  {journal} {Jetp Lett.}\ }\textbf {\bibinfo {volume} {70}},\ \bibinfo {pages}
  {609} (\bibinfo {year} {1999})}\BibitemShut {NoStop}%
\bibitem [{\citenamefont {Ivanov}(2001)}]{Ivanov2001}%
  \BibitemOpen
  \bibfield  {author} {\bibinfo {author} {\bibfnamefont {D.~A.}\ \bibnamefont
  {Ivanov}},\ }\bibfield  {title} {\bibinfo {title} {Non-abelian statistics of
  half-quantum vortices in $\mathit{p}$-wave superconductors},\ }\href
  {https://doi.org/10.1103/PhysRevLett.86.268} {\bibfield  {journal} {\bibinfo
  {journal} {Phys. Rev. Lett.}\ }\textbf {\bibinfo {volume} {86}},\ \bibinfo
  {pages} {268} (\bibinfo {year} {2001})}\BibitemShut {NoStop}%
\bibitem [{\citenamefont {Akzyanov}\ \emph {et~al.}(2015)\citenamefont
  {Akzyanov}, \citenamefont {Rakhmanov}, \citenamefont {Rozhkov},\ and\
  \citenamefont {Nori}}]{Akzyanov2015}%
  \BibitemOpen
  \bibfield  {author} {\bibinfo {author} {\bibfnamefont {R.~S.}\ \bibnamefont
  {Akzyanov}}, \bibinfo {author} {\bibfnamefont {A.~L.}\ \bibnamefont
  {Rakhmanov}}, \bibinfo {author} {\bibfnamefont {A.~V.}\ \bibnamefont
  {Rozhkov}},\ and\ \bibinfo {author} {\bibfnamefont {F.}~\bibnamefont
  {Nori}},\ }\bibfield  {title} {\bibinfo {title} {Majorana fermions at the
  edge of superconducting islands},\ }\href
  {https://link.aps.org/doi/10.1103/PhysRevB.92.075432} {\bibfield  {journal}
  {\bibinfo  {journal} {Phys. Rev. B}\ }\textbf {\bibinfo {volume} {92}},\
  \bibinfo {pages} {075432} (\bibinfo {year} {2015})}\BibitemShut {NoStop}%
\bibitem [{\citenamefont {Akzyanov}\ \emph
  {et~al.}(2016{\natexlab{a}})\citenamefont {Akzyanov}, \citenamefont
  {Rakhmanov}, \citenamefont {Rozhkov},\ and\ \citenamefont
  {Nori}}]{Akzyanov2016}%
  \BibitemOpen
  \bibfield  {author} {\bibinfo {author} {\bibfnamefont {R.~S.}\ \bibnamefont
  {Akzyanov}}, \bibinfo {author} {\bibfnamefont {A.~L.}\ \bibnamefont
  {Rakhmanov}}, \bibinfo {author} {\bibfnamefont {A.~V.}\ \bibnamefont
  {Rozhkov}},\ and\ \bibinfo {author} {\bibfnamefont {F.}~\bibnamefont
  {Nori}},\ }\bibfield  {title} {\bibinfo {title} {Tunable majorana fermion
  from landau quantization in 2d topological superconductors},\ }\href
  {https://link.aps.org/doi/10.1103/PhysRevB.94.125428} {\bibfield  {journal}
  {\bibinfo  {journal} {Phys. Rev. B}\ }\textbf {\bibinfo {volume} {94}},\
  \bibinfo {pages} {125428} (\bibinfo {year} {2016}{\natexlab{a}})}\BibitemShut
  {NoStop}%
\bibitem [{\citenamefont {Chiu}\ \emph {et~al.}(2016)\citenamefont {Chiu},
  \citenamefont {Teo}, \citenamefont {Schnyder},\ and\ \citenamefont
  {Ryu}}]{Chiu2016}%
  \BibitemOpen
  \bibfield  {author} {\bibinfo {author} {\bibfnamefont {C.-K.}\ \bibnamefont
  {Chiu}}, \bibinfo {author} {\bibfnamefont {J.~C.~Y.}\ \bibnamefont {Teo}},
  \bibinfo {author} {\bibfnamefont {A.~P.}\ \bibnamefont {Schnyder}},\ and\
  \bibinfo {author} {\bibfnamefont {S.}~\bibnamefont {Ryu}},\ }\bibfield
  {title} {\bibinfo {title} {Classification of topological quantum matter with
  symmetries},\ }\href {https://doi.org/10.1103/RevModPhys.88.035005}
  {\bibfield  {journal} {\bibinfo  {journal} {Rev. Mod. Phys.}\ }\textbf
  {\bibinfo {volume} {88}},\ \bibinfo {pages} {035005} (\bibinfo {year}
  {2016})}\BibitemShut {NoStop}%
\bibitem [{\citenamefont {Schnyder}\ \emph {et~al.}(2008)\citenamefont
  {Schnyder}, \citenamefont {Ryu}, \citenamefont {Furusaki},\ and\
  \citenamefont {Ludwig}}]{Schnyder2008}%
  \BibitemOpen
  \bibfield  {author} {\bibinfo {author} {\bibfnamefont {A.~P.}\ \bibnamefont
  {Schnyder}}, \bibinfo {author} {\bibfnamefont {S.}~\bibnamefont {Ryu}},
  \bibinfo {author} {\bibfnamefont {A.}~\bibnamefont {Furusaki}},\ and\
  \bibinfo {author} {\bibfnamefont {A.~W.~W.}\ \bibnamefont {Ludwig}},\
  }\bibfield  {title} {\bibinfo {title} {Classification of topological
  insulators and superconductors in three spatial dimensions},\ }\href
  {https://link.aps.org/doi/10.1103/PhysRevB.78.195125} {\bibfield  {journal}
  {\bibinfo  {journal} {Phys. Rev. B}\ }\textbf {\bibinfo {volume} {78}},\
  \bibinfo {pages} {195125} (\bibinfo {year} {2008})}\BibitemShut {NoStop}%
\bibitem [{\citenamefont {Volovik}(2003)}]{Volovik_book}%
  \BibitemOpen
  \bibfield  {author} {\bibinfo {author} {\bibfnamefont {G.~E.}\ \bibnamefont
  {Volovik}},\ }\href@noop {} {\emph {\bibinfo {title} {The Universe in a
  Helium Droplet}}}\ (\bibinfo  {publisher} {Clarendon, Oxford},\ \bibinfo
  {year} {2003})\BibitemShut {NoStop}%
\bibitem [{\citenamefont {Korhonen}\ \emph {et~al.}(1993)\citenamefont
  {Korhonen}, \citenamefont {Kondo}, \citenamefont {Krusius}, \citenamefont
  {Thuneberg},\ and\ \citenamefont {Volovik}}]{Korhonen1993}%
  \BibitemOpen
  \bibfield  {author} {\bibinfo {author} {\bibfnamefont {J.~S.}\ \bibnamefont
  {Korhonen}}, \bibinfo {author} {\bibfnamefont {Y.}~\bibnamefont {Kondo}},
  \bibinfo {author} {\bibfnamefont {M.}~\bibnamefont {Krusius}}, \bibinfo
  {author} {\bibfnamefont {E.~V.}\ \bibnamefont {Thuneberg}},\ and\ \bibinfo
  {author} {\bibfnamefont {G.~E.}\ \bibnamefont {Volovik}},\ }\bibfield
  {title} {\bibinfo {title} {Observation of combined spin-mass vortices in
  rotating $^{3}\mathit{B}$},\ }\href
  {https://doi.org/10.1103/PhysRevB.47.8868} {\bibfield  {journal} {\bibinfo
  {journal} {Phys. Rev. B}\ }\textbf {\bibinfo {volume} {47}},\ \bibinfo
  {pages} {8868} (\bibinfo {year} {1993})}\BibitemShut {NoStop}%
\bibitem [{\citenamefont {Wu}\ and\ \citenamefont {Martin}(2017)}]{Wu2017}%
  \BibitemOpen
  \bibfield  {author} {\bibinfo {author} {\bibfnamefont {F.}~\bibnamefont
  {Wu}}\ and\ \bibinfo {author} {\bibfnamefont {I.}~\bibnamefont {Martin}},\
  }\bibfield  {title} {\bibinfo {title} {Majorana kramers pair in a nematic
  vortex},\ }\href {https://doi.org/10.1103/PhysRevB.95.224503} {\bibfield
  {journal} {\bibinfo  {journal} {Phys. Rev. B}\ }\textbf {\bibinfo {volume}
  {95}},\ \bibinfo {pages} {224503} (\bibinfo {year} {2017})}\BibitemShut
  {NoStop}%
\bibitem [{\citenamefont {Venderbos}\ \emph
  {et~al.}(2016{\natexlab{a}})\citenamefont {Venderbos}, \citenamefont
  {Kozii},\ and\ \citenamefont {Fu}}]{Venderbos2016}%
  \BibitemOpen
  \bibfield  {author} {\bibinfo {author} {\bibfnamefont {J.~W.~F.}\
  \bibnamefont {Venderbos}}, \bibinfo {author} {\bibfnamefont {V.}~\bibnamefont
  {Kozii}},\ and\ \bibinfo {author} {\bibfnamefont {L.}~\bibnamefont {Fu}},\
  }\bibfield  {title} {\bibinfo {title} {Identification of nematic
  superconductivity from the upper critical field},\ }\href
  {https://doi.org/10.1103/PhysRevB.94.094522} {\bibfield  {journal} {\bibinfo
  {journal} {Phys. Rev. B}\ }\textbf {\bibinfo {volume} {94}},\ \bibinfo
  {pages} {094522} (\bibinfo {year} {2016}{\natexlab{a}})}\BibitemShut
  {NoStop}%
\bibitem [{\citenamefont {Kuntsevich}\ \emph {et~al.}(2018)\citenamefont
  {Kuntsevich}, \citenamefont {Bryzgalov}, \citenamefont {Prudkoglyad},
  \citenamefont {Martovitskii}, \citenamefont {Selivanov},\ and\ \citenamefont
  {Chizhevskii}}]{Kuntsevich2018}%
  \BibitemOpen
  \bibfield  {author} {\bibinfo {author} {\bibfnamefont {A.~Y.}\ \bibnamefont
  {Kuntsevich}}, \bibinfo {author} {\bibfnamefont {M.~A.}\ \bibnamefont
  {Bryzgalov}}, \bibinfo {author} {\bibfnamefont {V.~A.}\ \bibnamefont
  {Prudkoglyad}}, \bibinfo {author} {\bibfnamefont {V.~P.}\ \bibnamefont
  {Martovitskii}}, \bibinfo {author} {\bibfnamefont {Y.~G.}\ \bibnamefont
  {Selivanov}},\ and\ \bibinfo {author} {\bibfnamefont {E.~G.}\ \bibnamefont
  {Chizhevskii}},\ }\bibfield  {title} {\bibinfo {title} {Structural distortion
  behind the nematic superconductivity in sr x bi2se3},\ }\href
  {https://doi.org/10.1088/1367-2630/aae595} {\bibfield  {journal} {\bibinfo
  {journal} {New Journal of Physics}\ }\textbf {\bibinfo {volume} {20}},\
  \bibinfo {pages} {103022} (\bibinfo {year} {2018})}\BibitemShut {NoStop}%
\bibitem [{\citenamefont {Kuntsevich}\ \emph {et~al.}(2019)\citenamefont
  {Kuntsevich}, \citenamefont {Bryzgalov}, \citenamefont {Akzyanov},
  \citenamefont {Martovitskii}, \citenamefont {Rakhmanov},\ and\ \citenamefont
  {Selivanov}}]{Kuntsevich2019}%
  \BibitemOpen
  \bibfield  {author} {\bibinfo {author} {\bibfnamefont {A.~Y.}\ \bibnamefont
  {Kuntsevich}}, \bibinfo {author} {\bibfnamefont {M.~A.}\ \bibnamefont
  {Bryzgalov}}, \bibinfo {author} {\bibfnamefont {R.~S.}\ \bibnamefont
  {Akzyanov}}, \bibinfo {author} {\bibfnamefont {V.~P.}\ \bibnamefont
  {Martovitskii}}, \bibinfo {author} {\bibfnamefont {A.~L.}\ \bibnamefont
  {Rakhmanov}},\ and\ \bibinfo {author} {\bibfnamefont {Y.~G.}\ \bibnamefont
  {Selivanov}},\ }\bibfield  {title} {\bibinfo {title} {Strain-driven
  nematicity of odd-parity superconductivity in
  ${\mathrm{sr}}_{x}{\mathrm{bi}}_{2}{\mathrm{se}}_{3}$},\ }\href
  {https://doi.org/10.1103/PhysRevB.100.224509} {\bibfield  {journal} {\bibinfo
   {journal} {Phys. Rev. B}\ }\textbf {\bibinfo {volume} {100}},\ \bibinfo
  {pages} {224509} (\bibinfo {year} {2019})}\BibitemShut {NoStop}%
\bibitem [{\citenamefont {Abrikosov}(1957)}]{Abrikosov1957}%
  \BibitemOpen
  \bibfield  {author} {\bibinfo {author} {\bibfnamefont {A.~A.}\ \bibnamefont
  {Abrikosov}},\ }\bibfield  {title} {\bibinfo {title} {On the magnetic
  properties of superconductors of the second group},\ }\href
  {http://www.jetp.ac.ru/cgi-bin/e/index/e/5/6/p1174?a=list} {\bibfield
  {journal} {\bibinfo  {journal} {JETP}\ }\textbf {\bibinfo {volume} {5}},\
  \bibinfo {pages} {1442} (\bibinfo {year} {1957})}\BibitemShut {NoStop}%
\bibitem [{\citenamefont {Landau}\ and\ \citenamefont
  {Lifshitz}(1965)}]{LandauLifshtiz7}%
  \BibitemOpen
  \bibfield  {author} {\bibinfo {author} {\bibfnamefont {L.~D.}\ \bibnamefont
  {Landau}}\ and\ \bibinfo {author} {\bibfnamefont {E.~M.}\ \bibnamefont
  {Lifshitz}},\ }\href@noop {} {\emph {\bibinfo {title} {Theory of
  elasticity}}}\ (\bibinfo  {publisher} {Pergamon Press, Oxford},\ \bibinfo
  {year} {1965})\BibitemShut {NoStop}%
\bibitem [{\citenamefont {Fu}\ and\ \citenamefont {Berg}(2010)}]{Fu2010}%
  \BibitemOpen
  \bibfield  {author} {\bibinfo {author} {\bibfnamefont {L.}~\bibnamefont
  {Fu}}\ and\ \bibinfo {author} {\bibfnamefont {E.}~\bibnamefont {Berg}},\
  }\bibfield  {title} {\bibinfo {title} {Odd-parity topological
  superconductors: Theory and application to
  ${\mathrm{cu}}_{x}{\mathrm{bi}}_{2}{\mathrm{se}}_{3}$},\ }\href
  {https://doi.org/10.1103/PhysRevLett.105.097001} {\bibfield  {journal}
  {\bibinfo  {journal} {Phys. Rev. Lett.}\ }\textbf {\bibinfo {volume} {105}},\
  \bibinfo {pages} {097001} (\bibinfo {year} {2010})}\BibitemShut {NoStop}%
\bibitem [{\citenamefont {Yip}(2013)}]{Yip2013}%
  \BibitemOpen
  \bibfield  {author} {\bibinfo {author} {\bibfnamefont {S.-K.}\ \bibnamefont
  {Yip}},\ }\bibfield  {title} {\bibinfo {title} {Models of superconducting
  cu:bi${}_{2}$se${}_{3}$: Single- versus two-band description},\ }\href
  {https://doi.org/10.1103/PhysRevB.87.104505} {\bibfield  {journal} {\bibinfo
  {journal} {Phys. Rev. B}\ }\textbf {\bibinfo {volume} {87}},\ \bibinfo
  {pages} {104505} (\bibinfo {year} {2013})}\BibitemShut {NoStop}%
\bibitem [{\citenamefont {Venderbos}\ \emph
  {et~al.}(2016{\natexlab{b}})\citenamefont {Venderbos}, \citenamefont
  {Kozii},\ and\ \citenamefont {Fu}}]{Venderbos2016a}%
  \BibitemOpen
  \bibfield  {author} {\bibinfo {author} {\bibfnamefont {J.~W.~F.}\
  \bibnamefont {Venderbos}}, \bibinfo {author} {\bibfnamefont {V.}~\bibnamefont
  {Kozii}},\ and\ \bibinfo {author} {\bibfnamefont {L.}~\bibnamefont {Fu}},\
  }\bibfield  {title} {\bibinfo {title} {Odd-parity superconductors with
  two-component order parameters: Nematic and chiral, full gap, and majorana
  node},\ }\href {https://doi.org/10.1103/PhysRevB.94.180504} {\bibfield
  {journal} {\bibinfo  {journal} {Phys. Rev. B}\ }\textbf {\bibinfo {volume}
  {94}},\ \bibinfo {pages} {180504} (\bibinfo {year}
  {2016}{\natexlab{b}})}\BibitemShut {NoStop}%
\bibitem [{\citenamefont {Brems}\ \emph {et~al.}(2018)\citenamefont {Brems},
  \citenamefont {Paaske}, \citenamefont {Lunde},\ and\ \citenamefont
  {Willatzen}}]{Brems2018}%
  \BibitemOpen
  \bibfield  {author} {\bibinfo {author} {\bibfnamefont {M.~R.}\ \bibnamefont
  {Brems}}, \bibinfo {author} {\bibfnamefont {J.}~\bibnamefont {Paaske}},
  \bibinfo {author} {\bibfnamefont {A.~M.}\ \bibnamefont {Lunde}},\ and\
  \bibinfo {author} {\bibfnamefont {M.}~\bibnamefont {Willatzen}},\ }\bibfield
  {title} {\bibinfo {title} {Symmetry analysis of strain, electric and magnetic
  fields in the bi2se3-class of topological insulators},\ }\href
  {http://dx.doi.org/10.1088/1367-2630/aabcfc} {\bibfield  {journal} {\bibinfo
  {journal} {New Journal of Physics}\ }\textbf {\bibinfo {volume} {20}},\
  \bibinfo {pages} {053041} (\bibinfo {year} {2018})}\BibitemShut {NoStop}%
\bibitem [{\citenamefont {Deng}\ \emph {et~al.}(2020)\citenamefont {Deng},
  \citenamefont {Bonesteel},\ and\ \citenamefont {Schlottmann}}]{Deng2020}%
  \BibitemOpen
  \bibfield  {author} {\bibinfo {author} {\bibfnamefont {H.}~\bibnamefont
  {Deng}}, \bibinfo {author} {\bibfnamefont {N.}~\bibnamefont {Bonesteel}},\
  and\ \bibinfo {author} {\bibfnamefont {P.}~\bibnamefont {Schlottmann}},\
  }\bibfield  {title} {\bibinfo {title} {Bound fermion states in pinned
  vortices in the surface states of a superconducting topological insulator},\
  }\href {https://doi.org/10.1088/1361-648x/abba89} {\bibfield  {journal}
  {\bibinfo  {journal} {Journal of Physics: Condensed Matter}\ }\textbf
  {\bibinfo {volume} {33}},\ \bibinfo {pages} {035604} (\bibinfo {year}
  {2020})}\BibitemShut {NoStop}%
\bibitem [{\citenamefont {Akzyanov}\ \emph
  {et~al.}(2016{\natexlab{b}})\citenamefont {Akzyanov}, \citenamefont
  {Rakhmanov}, \citenamefont {Rozhkov},\ and\ \citenamefont
  {Nori}}]{Akzyanov2016a}%
  \BibitemOpen
  \bibfield  {author} {\bibinfo {author} {\bibfnamefont {R.~S.}\ \bibnamefont
  {Akzyanov}}, \bibinfo {author} {\bibfnamefont {A.~L.}\ \bibnamefont
  {Rakhmanov}}, \bibinfo {author} {\bibfnamefont {A.~V.}\ \bibnamefont
  {Rozhkov}},\ and\ \bibinfo {author} {\bibfnamefont {F.}~\bibnamefont
  {Nori}},\ }\bibfield  {title} {\bibinfo {title} {Tunable majorana fermion
  from landau quantization in 2d topological superconductors},\ }\href
  {https://doi.org/10.1103/PhysRevB.94.125428} {\bibfield  {journal} {\bibinfo
  {journal} {Phys. Rev. B}\ }\textbf {\bibinfo {volume} {94}},\ \bibinfo
  {pages} {125428} (\bibinfo {year} {2016}{\natexlab{b}})}\BibitemShut
  {NoStop}%
\end{thebibliography}%

\end{document}